\documentclass[%
 reprint,
superscriptaddress,
 amsmath,amssymb,
 aps,
pra,
]{revtex4-2}

\usepackage{upgreek}
\usepackage{color}

\usepackage{graphicx}
\usepackage{dcolumn}
\usepackage{bm}
\usepackage{hyperref}
\usepackage[normalem]{ulem}

\newcommand{\add}[1]{#1}

\begin{document}

\preprint{APS/123-QED}

\title{Squeezed light from a nanophotonic molecule}

\author{Y. Zhang}
\affiliation{Xanadu, Toronto, ON, M5G 2C8, Canada}
\author{M. Menotti}
\affiliation{Xanadu, Toronto, ON, M5G 2C8, Canada}
\author{K. Tan}
\affiliation{Xanadu, Toronto, ON, M5G 2C8, Canada}
\author{V.D. Vaidya}
\affiliation{Xanadu, Toronto, ON, M5G 2C8, Canada}
\author{D.H. Mahler}
\affiliation{Xanadu, Toronto, ON, M5G 2C8, Canada}
\author{\add{L.G. Helt}}
\affiliation{Xanadu, Toronto, ON, M5G 2C8, Canada}
\author{L. Zatti}
\affiliation{Department of Physics, University of Pavia, Via Bassi 6, 27100, Pavia, Italy}
\author{M. Liscidini}
\affiliation{Department of Physics, University of Pavia, Via Bassi 6, 27100, Pavia, Italy}
\author{B. Morrison}
\affiliation{Xanadu, Toronto, ON, M5G 2C8, Canada}
\author{Z. Vernon} \email{zach@xanadu.ai}
\affiliation{Xanadu, Toronto, ON, M5G 2C8, Canada}

\date{\today}

\maketitle

\noindent \textbf{Photonic molecules are composed of two or more optical resonators, arranged such that some of the modes of each resonator are coupled to those of the other. Such structures have been used \add{for emulating the behaviour of two-level systems \cite{spreeuw1990classical},} lasing \cite{grossmann2013polymeric}, and on-demand optical storage and retrieval \cite{zhang2019electronically}. Coupled resonators have also been used for dispersion engineering of integrated devices, enhancing their performance for nonlinear optical applications \cite{gentry2014tunable,heuck2019unidirectional}. Delicate engineering of such integrated nonlinear structures is required for developing scalable sources of non-classical light to be deployed in quantum information processing systems. In this work, we demonstrate a photonic molecule composed of two coupled microring resonators on an integrated nanophotonic chip, designed to generate strongly squeezed light uncontaminated by noise from unwanted parasitic nonlinear processes. By tuning the photonic molecule to selectively couple and thus hybridize only the modes involved in the unwanted processes, suppression of parasitic parametric fluorescence is accomplished. This strategy enables the use of microring resonators for the efficient generation of degenerate squeezed light: without it, simple single-resonator structures cannot avoid contamination from nonlinear noise without significantly compromising pump power efficiency, and are thus limited to generating only weak degenerate squeezing. We use this device to generate $\mathbf{8(1)}$ dB of \add{broadband} degenerate squeezed light on-chip, \add{with $\mathbf{1.65(1)}$ dB directly measured,}  which is the largest amount of squeezing yet reported from any nanophotonic source.}

Squeezed light sources \cite{lvovsky2015squeezed} are a fundamental building block of photonic technologies for quantum information processing. Squeezing is an essential resource for quantum sensing \cite{tse2019quantum} and a wide range of quantum computing algorithms \cite{huh2015boson,arrazola2018using,bromley2020applications,weedbrook2012gaussian}.  Recently, much effort has gone into engineering scalable implementations of such sources using integrated photonics. \add{Compact and highly efficient sources of squeezing have been developed based on whispering gallery mode resonators \cite{otterpohl2019squeezed}.} \add{Much progress has been made in developing squeezed light sources based on periodically poled waveguides in low-index-contrast platforms \cite{anderson1995quadrature,mondain2019chip}. But for applications requiring very large numbers of components, } high-index-contrast nanophotonic platforms are preferred, as they enable large-scale integration with many hundreds or thousands of \add{elements} on a single monolithic chip \cite{qiang2018large}. Within the domain of nanophotonic structures, bright intensity-difference squeezing in a silicon nitride ring resonator driven above the parametric oscillation threshold has been demonstrated \cite{dutt2015chip}. \add{Subsequently, a two-ring structure was used to enable tuning of the level of squeezing by varying the effective resonator coupling condition \cite{dutt2016tunable}.}  Single-ring resonators driven below threshold have yielded \emph{two-mode} (nondegenerate) quadrature squeezing and photon number difference squeezing \cite{vaidya2020broadband}. \add{Micromechanical resonators have generated small levels of squeezing over a few MHz of bandwidth \cite{safavi2013squeezed}.} Some squeezing in a single-mode degenerate configuration has also been reported with microrings using a single pump \cite{cernansky2019nanophotonic}, but this suffers from strong excess noise contributions arising from non-parametric effects such as thermorefractive fluctuation \cite{huang2019thermorefractive}. Aside from limiting the amount of available squeezing, the presence of excess noise is especially undesirable for quantum computing applications, as such noise degrades the purity of the quantum states employed. 

No nanophotonic device has yet been demonstrated which efficiently produces quadrature squeezed vacuum in a single, degenerate spectral mode, uncontaminated by excess noise from both non-parametric and unwanted parametric processes. A promising candidate has been proposed based on dual-pump spontaneous four-wave mixing in microring resonators \cite{guo2014telecom,vernon2019scalable}: two classical pumps tuned to independent resonances can produce squeezing in a single, degenerate spectral mode. \add{Such resonators are also desirable for the broadband nature of the squeezing they can produce \cite{ast2013high}.} By using pumps very well separated in frequency from the squeezed mode, noise contributions from non-parametric effects like thermorefractive fluctuations can be avoided. However, in such a system a number of unwanted \emph{parametric} effects \cite{helt2017parasitic} add noise to the squeezing band, irreversibly corrupting the output. The primary culprit for such unwanted noise is single-pump parametric fluorescence \cite{azzini2012ultra} driven by each individual pump; further degradation is caused by Bragg-scattering four-wave mixing \cite{agha2012low}, which can transfer energy away from the squeezed mode. 

The role of such parasitic parametric effects in degenerate squeezers based on single-resonator integrated structures was recently clarified by Zhao \emph{et al.} \cite{zhao2020near}. A silicon nitride ring resonator was used to generate squeezing in a dual-pump configuration. Without suppression of parasitic effects, it was shown that only $0.8$ dB of squeezing would have been achievable. To overcome this, some suppression of parasitic processes was achieved by detuning the pumps from resonance, with $1.34$ dB of degenerate squeezing observed and $3.09$ dB of squeezing inferred on-chip. However, this strategy suffers from a significant trade-off between squeezing and pump power efficiency, as detuning the pumps reduces their resonance enhancement in the ring, compromising the efficiency of the desired squeezing process.

\begin{figure*}
    \centering
    \includegraphics[width=0.7\textwidth]{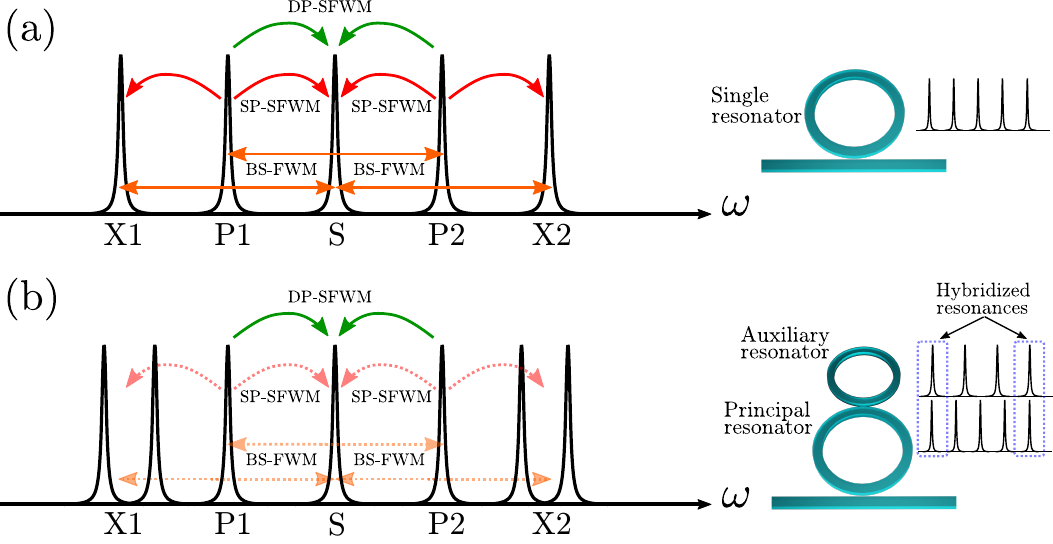}
    \caption{(a) Idealized intensity enhancement inside a single ring resonator, showing the four-wave mixing processes that occur when two resonances P1 and P2 are pumped. (b) Idealized intensity enhancement of the two-ring photonic molecule, showing the splitting and detuning of the hybridized X1 and X2 resonances that arises from the strong linear coupling between the principal and auxiliary resonator. In both cases the desired process, dual-pump spontaneous four-wave mixing (DP-SFWM, shown in green), leads to squeezing of the S mode. The unwanted processes of single-pump spontaneous four-wave mixing (SP-SFWM, red) generate excess noise in the S mode, contaminating the output, while Bragg-scattering four-wave mixing (BS-FWM, orange) transfers photons away from the S mode as photons are exchanged between the two pumps. Both these unwanted processes are suppressed by the presence of the auxiliary resonator, without significantly affecting the desired squeezing process. The free spectral ranges of the auxiliary resonator is chosen to be one third larger than that of the principal resonator, so that only every fourth mode of the principal resonator is hybridized.}
    \label{fig:schematic}
\end{figure*}

In this work, we bypass the detuning approach by directly engineering a device which strongly suppresses unwanted parasitic nonlinear effects without significantly compromising generation efficiency. Both single-pump parametric fluorescence and Bragg-scattering four-wave mixing effects are strongly enhanced by the presence of resonances that are otherwise not relevant to the desired dual-pump squeezing dynamics. These processes, and the associated resonances involved, are illustrated in Fig.~\ref{fig:schematic}(a). To suppress these unwanted processes, it suffices to design a structure for which the two resonances labelled X1 and X2 are removed or suitably corrupted, without significantly degrading the properties of the resonances used for the two pumps and the signal, labelled P1 and P2, and S, respectively. To that end, a photonic molecule based on a two-resonator structure was designed. As shown in Fig.~\ref{fig:schematic}(b), by tuning the auxiliary ring such that the X1 and X2 resonances of the principal ring nearly coincide in frequency with resonances of the auxiliary resonator, two new hybridized resonances emerge, strongly split and detuned from their original frequencies. The unwanted parametric processes involving the original X1 and X2 resonances are thereby highly suppressed. Since this modification to the X1 and X2 resonances can occur without having a significant impact on the P1, P2, and S resonances, strong enhancement of the desired squeezing process is maintained. This enables degenerate squeezing to be produced without degradation from unwanted single-pump parametric fluorescence and Bragg-scattering four-wave mixing, and without compromising the pump power efficiency of the source; more detail is available in the Supplemental Information. Similar multi-resonator schemes have been used to enable unidirectional frequency conversion \cite{heuck2019unidirectional} and dispersion compensation \cite{gentry2014tunable}. \add{In contrast to earlier work \cite{dutt2016tunable} that used a second ring resonator to tune the effective escape efficiency of a device for non-degenerate bright squeezing, our device enhances degenerate quadrature squeezing by \emph{selectively} suppressing unwanted processes.} 

The device was fabricated on a commercially available stoichiometric silicon nitride strip waveguide platform offered by Ligentec SA. The $\mathrm{Si}_3\mathrm{N}_4$ waveguide cross-section is $1500$~nm~$\times~800$~nm (width~$\times$~thickness) and is fully clad in $\mathrm{SiO}_2$. This platform and cross-section were selected for low propagation loss, lack of two-photon absorption, and high third-order optical nonlinearity. Independent microheaters were overlaid to provide thermal tuning of each resonator. The principal resonator was designed to have radius $R=114$~$\upmu\mathrm{m}$, and the auxiliary resonator to have radius $0.75\times R$, leading to free spectral ranges of $200$~GHz for the principal resonator, and $267$~GHz (approximately one third larger) for the auxiliary resonator. The principal resonator is strongly over-coupled to the bus waveguide, resulting in an escape efficiency of approximately $90\%$ in the wavelength range of interest; such over-coupling is important to limit the amount of loss experienced by the squeezed light on-chip.

A micrograph of the device and the measured linear transmission spectrum of the fundamental TE resonances in the wavelength range of interest are exhibited in Fig.~\ref{fig:transmission}(a) and Fig.~\ref{fig:transmission}(b), respectively. The auxiliary resonator microheater was tuned to achieve spectral alignment of the resonances associated with the auxiliary and principal resonator, leading to the formation of hybrid, split resonances associated with the unwanted single-pump parametric fluorescence and Bragg-scattering four-wave mixing. Three resonances of the principal resonator are preserved, displaying un-split Lorentzian lineshapes with loaded quality factors of approximately $3\times 10^5$. 

The transmission spectrum of the device for a range of different auxiliary microheater settings is plotted in Fig.~\ref{fig:transmission}(c) for wavelengths near the X1 resonance. As the auxiliary resonances are tuned, the resonance doublet exhibits the classic avoided crossing behaviour of coupled modes in a photonic molecule as the coupling strength (in this case determined by the detuning between X1 resonances of the auxiliary and principal resonators) is varied.

A simplified diagram of the experiment to generate and measure squeezing is shown in Fig.~\ref{fig:results}(a). Three phase- and frequency-locked beams were generated: two of these were amplified and filtered to serve as pumps, while the other beam was filtered and reserved for a use as a local oscillator (LO). The system used to generate the three beams --- electro-optic frequency comb beat note locking, followed by stimulated four-wave mixing in highly nonlinear fiber --- ensures that the local oscillator frequency is precisely equal to the average frequency of the pumps, and that the three optical phases are locked, as required for homodyne detection. One pump was tuned to the resonance near $1560.9$~nm (the P1 resonance), and the principal resonator was actively locked to this pump to maximize the circulating pump power in the principal ring. The frequency of the other pump was tuned to a second resonance near $1557.7$~nm (P2). The pump powers were set to be equal, and could be adjusted between $0$ and $100$ mW total power on-chip\add{; pump powers reported in this work refer to the sum power of both pumps}. The generated squeezed light was separated from the residual pumps and measured via balanced homodyne detection as the LO phase was ramped. Quadrature variances at sideband frequencies up to $1.0$ GHz were recorded on an electrical spectrum analyzer. More details are available in the Methods section and Supplemental Information.

A representative quadrature variance trace at $20$ MHz sideband frequency is shown in Fig.~\ref{fig:results}(b) as the LO phase is ramped. The directly measured squeezing was $1.65(1)$dB. As the total collection and detection efficiency was $38(2)\%$ (factoring in all losses experienced by the squeezed light except the resonator escape efficiency), this corresponds to approximately $8(1)$ dB of squeezing available at the device output on-chip---the largest amount of squeezing of any kind yet reported from a nanophotonic device. This is consistent with the maximum amount of squeezing possible from this device: as the principal resonator escape efficiency is approximately $90\%$, the maximum amount of squeezing on-chip is limited to $10$ dB. Higher escape efficiencies, achieved by increasing the intrinsic quality factor of the device or increasing the coupling between the principal resonator and the bus waveguide would enable squeezing well in excess of $10$ dB. \add{For comparison, the level of squeezing required for fault-tolerant continuous variable quantum computation was recently shown to be approximately $10$ dB \cite{fukui2018high}, and the record highest squeezing across all platforms is currently $15$ dB from a hemilithic bulk-optical cavity \cite{vahlbruch2016detection}.} \add{Finally, we estimate the quantum state purity $(V_\mathrm{min}V_\mathrm{max})^{-1/2}$ \cite{paris2003purity}, where $V_\mathrm{min}$ and $V_\mathrm{max}$ are respectively the directly measured minimum and maximum quadrature variances of the $20$ MHz sideband mode, to be approximately $0.7$.}

The measured squeezing and anti-squeezing at $20$ MHz sideband frequency as a function of pump power are shown in Fig.~\ref{fig:results}(c). Optimal squeezing was obtained with approximately $70$ mW of total on-chip pump power\add{, which is approximately half of the optical parametric oscillation (OPO) threshold.} Beyond that power, the squeezing degrades as the OPO threshold is approached\add{, possibly due to the finite phase stability of the LO and rapidly increasing variance of the anti-squeezed quadrature as the pump power is increased \cite{aoki2006squeezing}.}  The squeezing is broadband: The squeezing and anti-squeezing spectra are exhibited in Fig.~\ref{fig:results}(d), demonstrating that squeezing can be observed even for sideband frequencies up to $1$ GHz. This is limited by the resonator linewidth and detection bandwidth. 

\add{To benchmark our results, the measured squeezing and anti-squeezing power scaling and spectra were fit to a theoretical model for degenerate squeezing in microring resonators with unwanted processes suppressed. These fits are exhibited by the solid lines in Figs. \ref{fig:results}(c) and \ref{fig:results}(d), and show good agreement between theory and experiment. Independently measured values for the system loss, resonance linewidth, and pump detuning were used for these curves. In Fig. \ref{fig:results}(d), this leaves only the gain $g$, a dimensionless number proportional to the pump power and equal to unity at the OPO threshold, as the free parameter. This fit extracts $g\approx 0.46$. The free parameter in the power scaling plot \ref{fig:results}(c) is the proportionality constant that relates the gain $g$ and the pump power; this fit predicts a gain $g\approx 0.50$ at $70$ mW, the pump power that optimizes squeezing. These independently extracted fit parameters are mutually consistent, and are also approximately consistent with our estimates for the OPO threshold being between $130$ and $160$ mW. This provides important validation for our model of the device, and further underscores the effectiveness of noise suppression in this system. More detail on the theoretical model and fitting procedure can be found in the Supplemental Information.} 

To assess the importance of the auxiliary resonator in suppressing unwanted processes, the parametric fluorescence noise spectrum generated in the S mode was measured with only the pump at P1 turned on, with $70$ mW on-chip power. The results are shown in Fig.~\ref{fig:results}(e) for two different voltages applied to the auxiliary resonator. When the auxiliary resonator is tuned such that the X1 resonance is no longer hybridized, a strong noise contribution -- more than $1.2$~dB above shot noise -- on all quadratures is observed. This excess noise is reduced to less than $0.1$~dB above shot noise when the auxiliary resonator is tuned appropriately. In the absence of the auxiliary resonator, several dB of excess noise would therefore be present in the S mode, severely degrading the purity and achievable squeezing in the generated quantum state. \add{This effect can also be directly seen in Fig. \ref{fig:results}(f), in which the on-chip squeezing and anti-squeezing spectra are shown with noise suppression enabled (orange crosses) and disabled (blue points). For fair comparison, the power was adjusted from $70$ mW for the data with suppression enabled to $90$ mW for the data with suppression disabled, in order to maintain a fixed degree of anti-squeezing. This adjustment in power was necessary to compensate for the small perturbations in the effective quality factors and resonance frequencies of the principal ring that arise from tuning the auxiliary resonator.} 

We have demonstrated degenerate quadrature squeezed vacuum with a noise reduction of $1.65(1)$~dB below shot noise, corresponding to $8(1)$~dB of squeezing available on-chip. The suppression of unwanted parametric processes was crucial to demonstrate strong single-mode squeezed light sources based on four-wave mixing. This was made possible by a designing a nanophotonic molecule to selectively suppress unwanted parasitic processes without significantly affecting squeezing efficiency. These results highlight the significant control that can be achieved over quantum nonlinear optical processes by exploiting nanophotonic platforms, and remove a significant barrier impeding progress towards scaling up devices for photonic quantum information processing.

\section*{Methods}
\begin{figure*}
    \centering
    \includegraphics[width=0.95\textwidth]{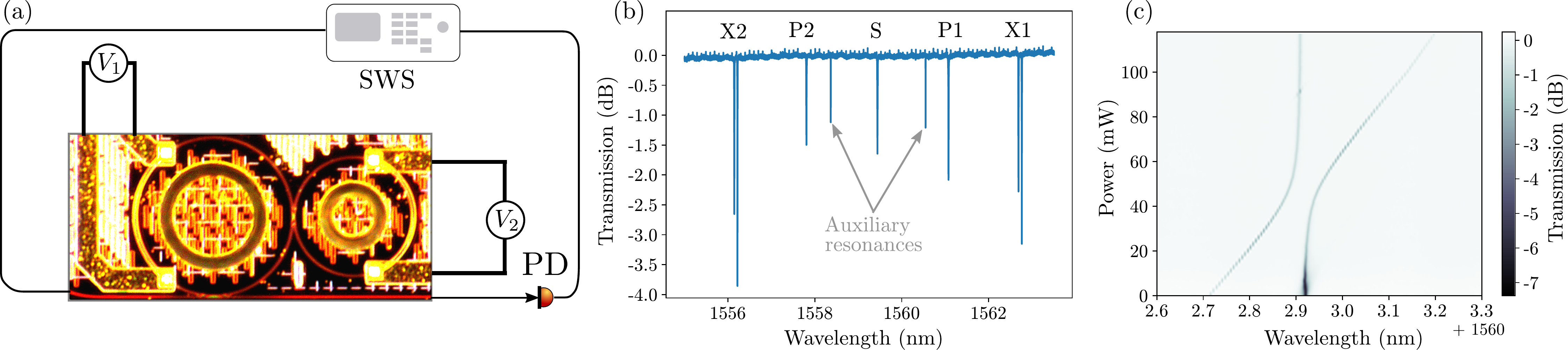}
    \caption{(a) Micrograph of the photonic molecule structure and simplified schematic of the apparatus for linear characterization. The principal resonator is on the left, coupled at the bottom to a bus waveguide. The smaller ring on the right acts as the auxiliary resonator. Microheaters overlaid apply voltages $V_1$ and $V_2$ to the principal and auxiliary heaters, respectively. A swept wavelength source (SWS) and photodiode (PD) measure the transmission spectrum of the device. (b) Measured TE polarization transmission spectrum of the device in the wavelength range of interest, with resonators tuned to hybridize the unwanted resonances X1 and X2.  Also evident are two resonances of the auxiliary resonator, which are indirectly weakly coupled to the bus waveguide via the principal resonator. (c) Transmission spectrum near the X1 resonance doublet of the device as the power dissipated by the auxiliary microheater is scanned. The resonance doublet exhibits the classic ``avoided crossing" behaviour associated with a pair of hybridized modes of a photonic molecule.}
    \label{fig:transmission}
\end{figure*}

The chip was fabricated on a dedicated wafer run using a commercially available photolithographic process offered by Ligentec SA. The cross-section of the bus waveguide was $1000$ nm $\times$ $800$ nm (width x height), which is tapered up to $1500$ nm width near the ring to match the its cross-section. While the desired dual-pump squeezing process would be optimized with normal dispersion, the ring waveguide cross-section resulted in weak anomalous dispersion; this was not an important factor, since the wavelength range used was relatively narrow. \add{While self- and cross-phase modulation result in a pump power-dependent change in the effective dispersion of the principal resonator, their effects did not significantly hinder the device performance: the induced detunings were less than a half-linewidth in magnitude, and their effects were easily compensated for by a modest increase in pump power.} \add{The microheater placement was designed to minimize thermal cross-talk. The resulting wavelength tuning cross-talk is approximately $15\%$, i.e., a wavelength shift of $d\lambda$ applied via the microheater on one ring results in a shift of about $0.15d\lambda$ in the other ring's resonance spectrum.} Edge couplers were used for coupling light in and out of the chip; ultrahigh numerical aperture (Nufern UHNA7) optical fibers were aligned to the chip facets, with index matching gel used to suppress reflections. The chip was temperature-stabilized using a thermoelectric cooler. \add{The reported escape efficiency was extracted by individual least-squares fitting of the pump and signal resonances to a model for a lossy ring resonator. This resulted in escape efficiencies of $90\%$ and $92\%$ for the two pumps, and $91\%$ for the signal resonance. Rounding to one significant figure, we report this as ``approximately $90\%$". This value is consistent with the extinction ratios of the resonances evident in Fig. \ref{fig:transmission}(b).}

\begin{figure*}
    \centering
    \includegraphics[width=0.82\textwidth]{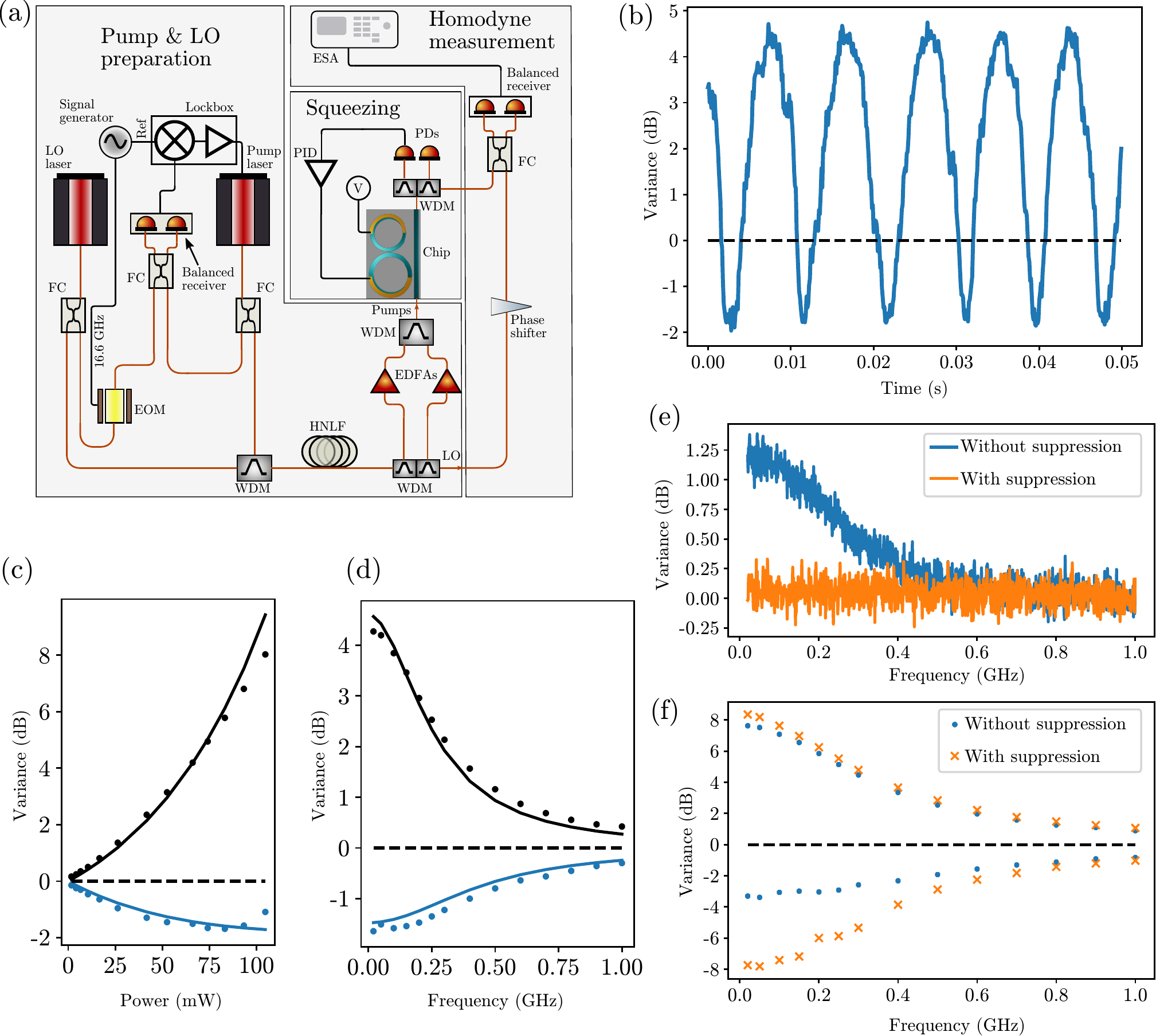}
    \caption{(a) Simplified diagram of experimental setup used for squeezing measurements. Two lasers are phase-locked by monitoring the beat note generated when interfering them after phase-modulating one with a $16.6$ GHz tone using a fast electro-optic modulator (EOM). These lasers serve as the local oscillator (LO) and first pump. They are then combined and coupled to a length of highly-nonlinear fiber (HNLF) to generate a second pump, automatically phase-locked to the first pump and LO, via stimulated four-wave mixing.   The two pumps are then separated from the LO, amplified by erbium-doped fiber amplifiers (EDFAs), and injected into the chip by edge couplers. The principal resonator of the photonic molecule is locked to the pump lasers' wavelengths by a slow proportional-integral-derivative (PID) loop, while the auxiliary resonator voltage supply (V) is set to maximize suppression of unwanted processes. Balanced homodyne detection is performed on the squeezed light output, as the LO phase is ramped. The quadrature variances are then recorded by an electrical spectrum analyzer (ESA). FC: fiber coupler; PD: photodiode; WDM: wavelength division multiplexing filter. More detail and a full experimental diagram are available in the Supplemental Information.
(b) Quadrature variance at $20$~MHz sideband frequency (blue trace) normalized to shot noise, plotted as a function of time as the LO phase is ramped. The black dashed line is the shot noise level.
(c) Measured squeezing (blue points) and anti-squeezing (black points) at $20$ MHz sideband frequency as the on-chip pump power is varied. Black dashed line is shot noise level; solid lines are a fit of the data points to a theoretical model. The three data points at the highest powers were omitted from the fit, as they are significantly affected by phase noise, which is not included in the model.
(d) The squeezing (blue points) and anti-squeezing (black points) spectra, taken at $70$ mW on-chip pump power. Black dashed line is shot noise level; Solid lines are a fit of the data points to a theoretical model.
(e) Single-pump parametric fluorescence spectrum (normalized to shot noise) in the S mode, measured using homodyne detection with one pump turned off. The blue trace is taken with the auxiliary resonator tuned such that the P1, S, and X1 resonances are not significantly coupled to the principal resonator, effectively disabling the noise suppression, resulting in the contamination of the signal mode with broadband unwanted noise. The orange trace is taken with the auxiliary resonator tuned to split the X1 and X2 resonances (Fig.~\ref{fig:transmission}(b)), suppressing the noise to less than $0.1$~dB above shot noise over the entire measurement band. \add{For both traces the pump power was $75$ mW, and the noise observed is phase-insensitive, as expected for single-pump parametric fluorescence.}
\add{(f) On-chip squeezing and anti-squeezing spectra with noise suppression enabled (orange crosses) and disabled (blue points). The pump power was adjusted from $70$ mW for the suppression-enabled case to $90$ mW for the suppression-disabled case to keep the anti-squeezing approximately fixed, compensating for small changes in the overall four-wave mixing efficiency associated with tuning the auxiliary resonator. Squeezing is strongly diminished with noise suppression disabled.} }
    \label{fig:results}
\end{figure*}

Active optical components in Fig.~\ref{fig:results}(a) included: Continuous wave tunable diode lasers (Pure Photonics PPCL 550), a fiber-coupled fast electro-optic phase modulator (EOSpace), $200$ m of highly nonlinear fiber (Sumitomo HNDS 1600BA-5), erbium-doped fiber amplifiers (Amonics AEDFA-33-B), and a fiber-coupled piezoelectric phase shifter (General Photonics FPS-001). Passive optical components included commercially available fiber-coupled taps, couplers, WDM filters, and polarization controllers. A multi-channel variable optical attenuator (OZ Optics) was used for controlling the powers of the pumps and LO. Optical isolators were placed before the nonlinear fiber, and before and after the chip to control back-reflection.

The electro-optic frequency comb locking used a commercially available lockbox (Vescent D2-135), and actuated the fast current input on one of the lasers. The $16.6$~GHz tone was generated with a microwave signal generator (Syntotic DS-3000). The system used to generate the pumps and LO result in the relevant phase parameter $2\phi_\mathrm{LO}-\phi_\mathrm{P1}-\phi_\mathrm{P2}$ being stabilized to within $2$ degrees root mean square, with $\phi_\mathrm{LO}$, $\phi_\mathrm{P1}$, and $\phi_\mathrm{P2}$ the phases of the LO, P1 and P2, respectively. The slow PID loop used to lock the principal resonator to the pump lasers employed a field programmable gate array board (Red Pitaya) running PyRPL \cite{neuhaus2017pyrpl}.

Homodyne detection was carried out using a tunable fiber-coupled splitter (Newport) and a commercially available balanced receiver (Wieserlabs WL-BPD1GA, another copy also used for beat note detection in the phase locking loop). The quantum efficiency of the detectors was approximately $80\%$, and the bandwidth $1$ GHz. The local oscillator power was set such that the detectors were operated well within the linear regime, with \add{14.7} dB of dark noise clearance \add{at 20 MHz sideband frequency, gradually declining to $12$ dB at 1 GHz; the full dark noise and shot noise spectra are available in the Supplementary Information}. An electrical spectrum analyzer (Keysight NA9020A) was used to measure the photocurrent difference fluctuations, with the resolution bandwidth set to $1$ MHz, video bandwidth $100$ Hz. On-chip squeezing was inferred from measured squeezing by the formula $V_\mathrm{on-chip}=(V_\mathrm{meas}+\eta-1)/\eta$, where $\eta$ is the total collection efficiency after the ring output (including chip outcoupling loss, fiber component loss, and detector quantum efficiency), $V_\mathrm{meas}$ the measured minimum quadrature variance (relative to vacuum), and $V_\mathrm{on-chip}$ the inferred quadrature variance on-chip. \add{The collection efficiency $\eta$ was deduced by directly measuring, using classical light, the chip outcoupling loss and the loss in the fiber components after the chip. The detector quantum efficiency was calculated from the photodiode responsivity quoted by the manufacturer. The experimental uncertainty of $0.02$ in $\eta$ is propagated through the calculation of the on-chip squeezing to arrive at a final uncertainty of $1$ dB in the $8$ dB estimate.} 

More details are available in the Supplemental Information. All data required to evaluate the conclusions of this work are available from the authors upon request.

\clearpage
\bibliography{references}

\widetext
\clearpage
\begin{center}
\textbf{\large Supplemental Information}
\end{center}
\setcounter{equation}{0}
\setcounter{figure}{0}
\setcounter{table}{0}
\setcounter{page}{1}
\makeatletter
\renewcommand{\theequation}{S\arabic{equation}}
\renewcommand{\thefigure}{S\arabic{figure}}
\renewcommand{\thesection}{S\arabic{section}}
\section{Squeezing experiment}
\begin{figure}[h]
    \centering
    \includegraphics[width=0.98\textwidth]{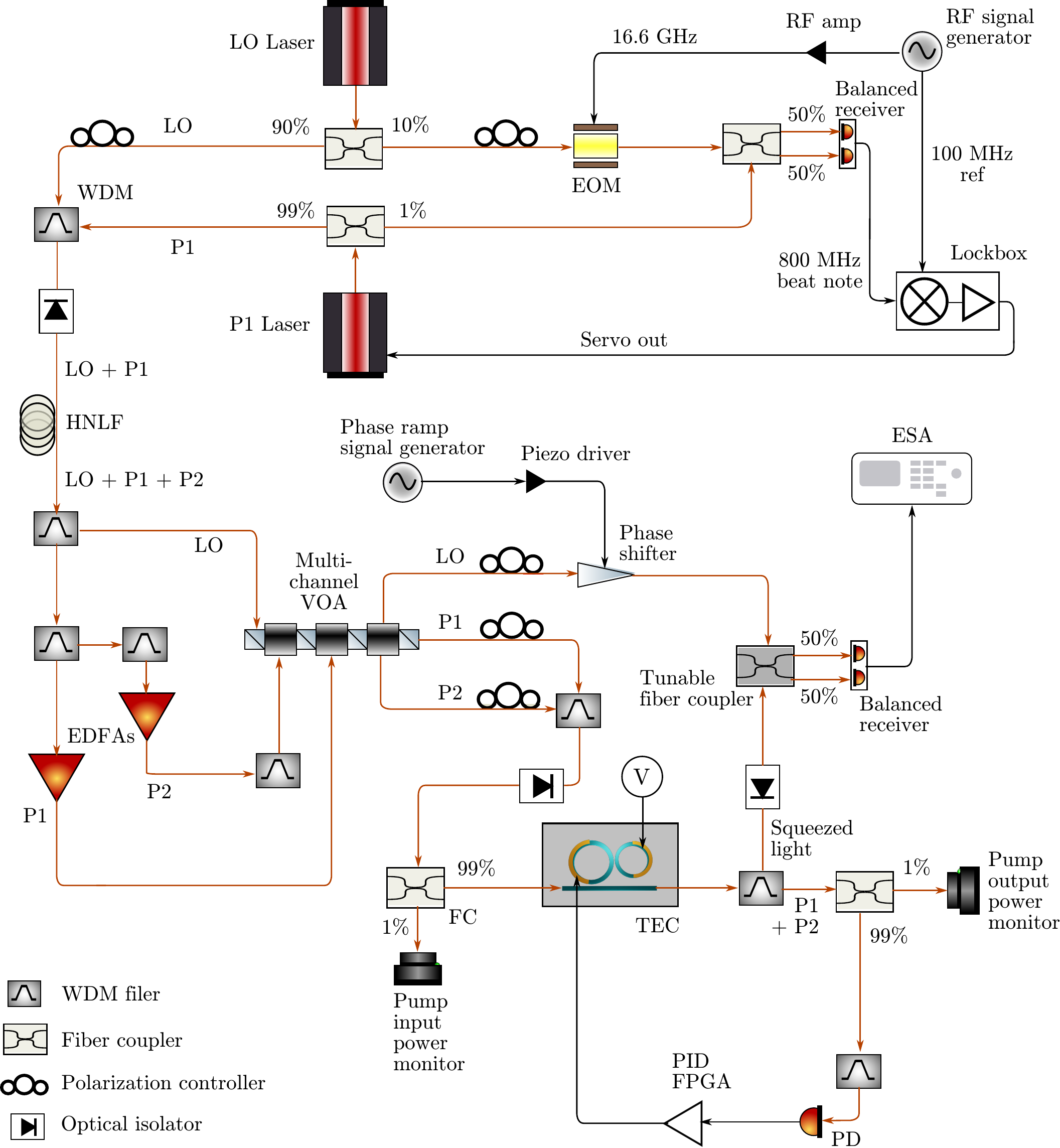}
    \caption{Full schematic of experimental setup for measuring squeezing. Details in text.}
    \label{fig:supp-experiment}
\end{figure}

A full diagram of the experimental setup used for measuring squeezing is shown in Fig.~\ref{fig:supp-experiment}. The apparatus can be divided into three functional blocks: Pump and local oscillator (LO) preparation, squeezing, and homodyne measurement. These are detailed in the following sections. 

\subsection{Pump and LO preparation}
Homodyne detection of squeezed light generated from dual-pump four-wave mixing requires a bright LO beam with a phase $\phi_\mathrm{LO}$ that is stable with respect to the sum phase $\phi_\mathrm{P1} +\phi_\mathrm{P2}$ of the pumps. That is, the phase parameter $2\phi_\mathrm{LO}-\phi_\mathrm{P1}-\phi_\mathrm{P2}$ must be stable, up to slow drifts that can be mitigated by a slow phase shifter. In addition, the LO frequency $f_\mathrm{LO}$ must precisely equal the average frequency $(f_\mathrm{P1}+f_\mathrm{P2})/2$ of the pumps. Finally, for stable operation, the frequency difference $f_\mathrm{P1} - f_\mathrm{P2}$ should be locked to a stable value.

To generate two pumps and an LO that satisfy these conditions, a hybrid radiofrequency (RF) electro-optic frequency comb locking and stimulated four-wave mixing scheme was employed. Two tunable diode lasers (Pure Photonics PPCL 550) generate the LO and first pump (P1) beams. A $10\%$ fraction of the LO beam is tapped, aligned in polarization, and deeply phase-modulated by a fast electro-optic phase modulator (EOSpace) driven by a strong $16.6$ GHz tone synthesized by RF signal generator (Syntotic DS-3000) and amplified by an RF amplifier (Minicircuits ZVE-3W-183+). The signal generator also outputs a $100$ MHz reference tone, which is directed to a lockbox (Vescent D2-135) that includes a mixer and proportional-integral-derivative (PID) controller. The phase-modulated LO light is interfered on a 50:50 fiber coupler with a $1\%$ fraction of the P1 light tapped off by a fiber coupler. The outputs of the 50:50 coupler are measured by a fiber-coupled balanced receiver (Wieserlabs WL-BPD1GA), with the electronic signal directed to the lockbox. In the lockbox, the $800$ MHz beat note between the P1 beam and the 12th-order sideband of the phase-modulated LO beam is mixed down and beat against the $100$ MHz reference, yielding a feedback signal directed to the fast current input of the P1 laser. This loop locks the optical frequencies of the LO and P1 lasers. While not necessary for this application, it also provides a degree of phase locking between the P1 pump and LO beams (to within approximately $12$ degrees root-mean-square). 

With the frequency locking loop persistently active, the remaining  P1 and LO light are aligned in polarization and combined by a wavelength division multiplexing (WDM) filter and passed through an optical isolator (General Photonics) to limit back-reflections. With about $40$ mW of power present at each wavelength, the light was then propagated through $200$ m of highly nonlinear optical fiber (HNLF, Sumitomo HNDS 1600BA-5). Stimulated four-wave mixing in this fiber generated a bright beam with frequency $f_\mathrm{P2}=2f_\mathrm{LO}-f_\mathrm{P1}$ for the second pump P2. This nonlinear process also ensures that the phase $2\phi_\mathrm{LO}-\phi_\mathrm{P1}-\phi_\mathrm{P2}$ is precisely fixed, as required. About $0.3$ mW of power at $f_\mathrm{P2}$ is generated and available at the end of the HNLF fiber spool. 

After the HNLF, the LO and two pumps are separated by WDM filters. The pumps are individually amplified by a pair of erbium-doped fiber amplifiers (Amonics AEDFA-33-B). The pumps and LO are sent through separate channels of a multi-channel digital variable optical attenuator (VOA, Oz Optics) for independent power control. Each of the three beams are independently aligned in polarization, with the LO then directed to a fiber-coupled piezoelectric phase shifter (General Photonics FPS-001) driven by a slow signal generator (Wavestation 2052) and piezo driver (Piezo Drive PD200). The pumps are combined by a WDM filter, passed through another optical isolator, with their total power monitored by a $1\%$ tap and slow photodiode. 

\subsection{Squeezing}
The two pumps are injected into the chip via ultra-high numerical aperture (Nufern UHNA7) fiber and a waveguide edge coupler. Index matching gel is used between the fiber and chip to suppress reflections at the facet. The chip is mounted on a thermo-electric cooler (TEC), which stabilizes the chip temperature via slow active feedback control carried out by PID loop controller (Arroyo 5240). A voltage supply (V) is used to set the auxiliary ring microheater voltage. The chip was wire bonded to ensure stable electrical control.

The squeezed light and residual pump light is coupled out of the chip to a WDM filter, which separates the two pumps from the squeezed beam. The two pump outputs are monitored by a slow photodiode, with the remaining pump light used as a signal to lock the principal ring to the P1 pump frequency. A photodiode actively monitors the P1 transmission, with the signal directed to a Red Pitaya field programmable gate array (FPGA) board running PyRPL \cite{neuhaus2017pyrpl}. This implements a PID loop actuating the principal ring microheater, locking the principal resonator to the pump frequency.

The total collection and detection efficiency experienced by the squeezed light after it exits the resonator is $38\%$: $71\%$ from the chip out-coupling, $67\%$ from the fiber components after the chip, and $80\%$ from the quantum efficiency of homodyne detector. We estimate our confidence in this to be within about $\pm 2\%$. Accounting for the $90\%$ resonator escape efficiency, the total system efficiency experienced by the squeezed light during and after generation is $34\%$. This is consistent with the degree of squeezing and anti-squeezing observed when the auxiliary resonator noise suppression is optimal.

\subsection{Homodyne measurement}
\begin{figure}
    \centering
    \includegraphics{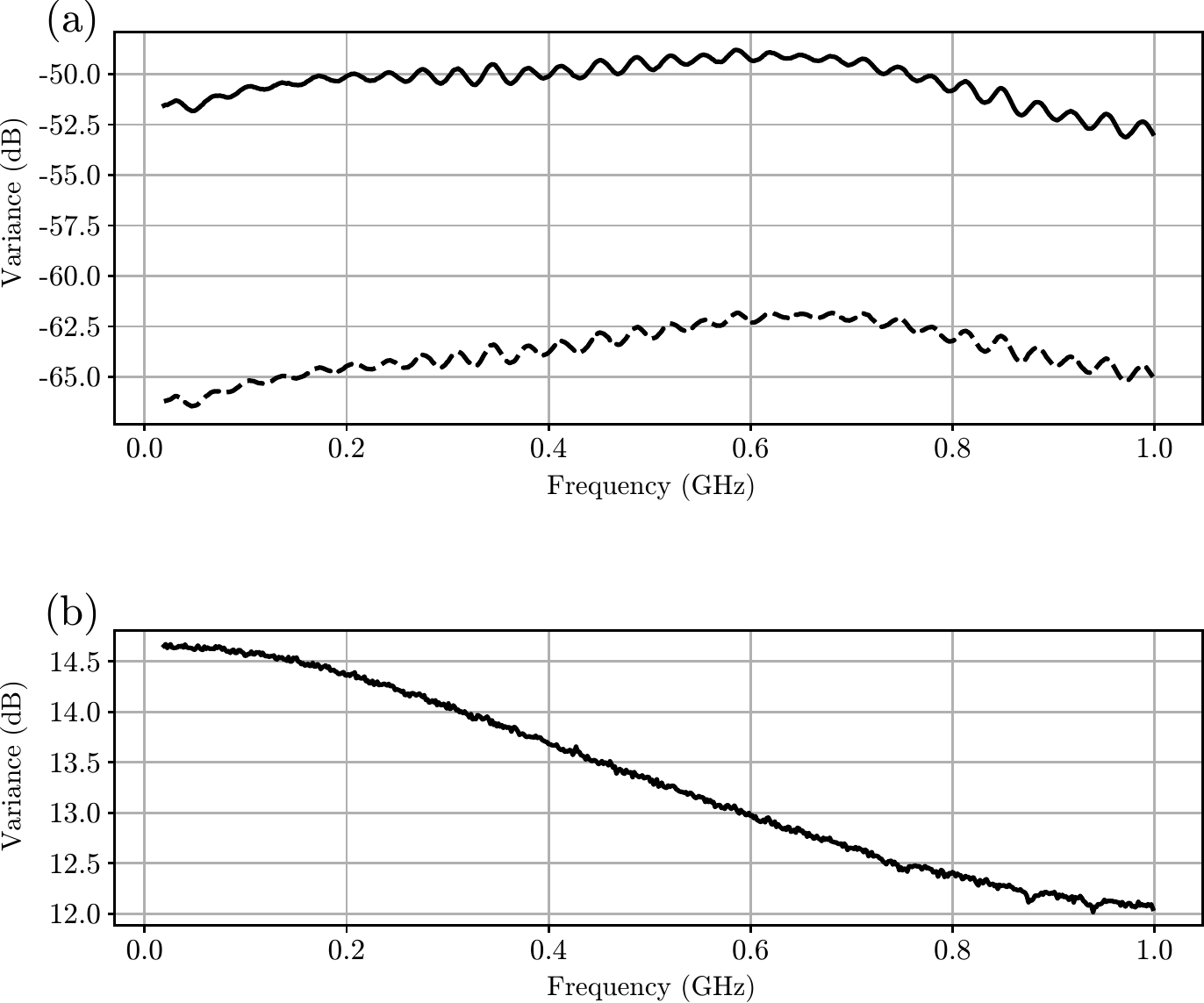}
    \caption{(a) Dark noise (dashed line) and shot noise (solid line) spectra from $20$ MHz to $1$ GHz sideband frequencies. The shot noise trace is acquired with $12$ mW LO power. (b) Noise clearance (shot noise normalized to dark noise) spectrum from $20$ MHz to $1$ GHz sideband frequencies for $12$ mW LO power. The clearance varies from $14.7$ dB at low frequencies to $12.0$ dB at high frequencies.}
    \label{fig:clearance}
\end{figure}

The squeezed light output from the chip is interfered with the LO on a tunable fiber coupler (Newport), which is adjusted to obtain a precise 50:50 splitting ratio. The LO phase was modulated with a $100$ Hz triangle wave signal. The two outputs from the tunable fiber coupler are incident on another balanced receiver (Wieserlabs WL-BPD1GA). The receiver has $1$ GHz bandwidth, $30$ dB common mode rejection ratio, $3500$ V/W transimpedance gain, and is AC coupled. The quantum efficiency of the detectors is approximately $80\%$. \add{The dark noise and shot noise spectra from $20$ MHz to $1$ GHz are shown in Fig. \ref{fig:clearance}(a); the dark noise clearance (i.e., shot noise normalized to dark noise) is plotted in Fig. \ref{fig:clearance}(b). At the $20$ MHz sideband frequency used for the squeezing data in Fig. \ref{fig:results}(b) and \ref{fig:results}(c), the clearance is $14.7$ dB; this gradually declines to $12.0$ dB at $1$ GHz. The LO power used was $12$ mW.}  The output signal from the receiver was connected to an electrical spectrum analyzer (ESA, Keysight NA9020A) with set to resolution bandwidth $1$ MHz, video bandwidth $100$ Hz.

\section{Suppression of parasitic processes}
The dual-pump scheme shown in Fig.~\ref{fig:schematic}(a) of the main text leads to a number of different processes associated with the $\chi^{(3)}$ nonlinear interaction. These include dual-pump spontaneous four-wave mixing (DP-SFWM), which is the desired interaction for producing degenerate squeezed vacuum states, single-pump SFWM (SP-SFWM), which generates photon pairs with one photon in mode S and one photon in either mode X1 or X2, and Bragg-scattering FWM (BS-FWM), which transfers a photon between S and X1 or X2 as a photon is transferred between P1 and P2. Finally, the third-order nonlinearity affects the spectral position of the resonant frequencies of the structure through self-phase modulation (SPM) and cross-phase modulation (XPM). The nonlinear Hamiltonian describing these processes can be written as

\begin{equation}
    H_{\mathrm{NL}}=H_{\mathrm{DP-SFWM}}+H_{\mathrm{SPM}}+H_{\mathrm{XPM}}+H_{\mathrm{SP-SFWM}}+H_{\mathrm{BS-FWM}},
\end{equation}
where
\begin{align}
H_\mathrm{\mathrm{DP-SFWM}}&=-\hbar\Lambda\left(b_\mathrm{S}^{\dagger}b_\mathrm{S}^{\dagger}b_\mathrm{P1}b_\mathrm{P2}+\mathrm{H.c.}\right),\\
H_\mathrm{\mathrm{SPM}}&=-\hbar\frac{\Lambda}{2}(b_\mathrm{P1}^{\dagger}b_\mathrm{P1}^{\dagger}b_\mathrm{P1}b_\mathrm{P1}+b_\mathrm{P2}^{\dagger}b_\mathrm{P2}^{\dagger}b_\mathrm{P2}b_\mathrm{P2}),\\
H_\mathrm{\mathrm{XPM}}&=-2\hbar\Lambda\left(b_\mathrm{S}^{\dagger}b_\mathrm{P1}^{\dagger}b_\mathrm{S}b_\mathrm{P1}+b_\mathrm{S}^{\dagger}b_\mathrm{P2}^{\dagger}b_\mathrm{S}b_\mathrm{P2}+b_\mathrm{P1}^{\dagger}b_\mathrm{P2}^{\dagger}b_\mathrm{P1}b_\mathrm{P2}+b_\mathrm{X1}^{\dagger}b_\mathrm{P1}^{\dagger}b_\mathrm{X1}b_\mathrm{P1}\right.\nonumber \\&+\left.b_\mathrm{X1}^{\dagger}b_\mathrm{P2}^{\dagger}b_\mathrm{X1}b_\mathrm{P2}+b_\mathrm{X2}^{\dagger}b_\mathrm{P1}^{\dagger}b_\mathrm{X2}b_\mathrm{P1}+b_\mathrm{X2}^{\dagger}b_\mathrm{P2}^{\dagger}b_\mathrm{X2}b_\mathrm{P2}\right),\\
H_\mathrm{\mathrm{SP-SFWM}}&=-\hbar\Lambda\left(b_\mathrm{X1}^{\dagger}b_\mathrm{S}^{\dagger}b_\mathrm{P1}b_\mathrm{P1}+b_\mathrm{X2}^{\dagger}b_\mathrm{S}^{\dagger}b_\mathrm{P2}b_\mathrm{P2}\right)+\mathrm{H.c.},\\
H_\mathrm{\mathrm{BS-FWM}}&=-2\hbar\Lambda\left( b_\mathrm{P1}b_\mathrm{P2}^{\dagger}b_\mathrm{S}b_\mathrm{X1}^{\dagger}+ b_\mathrm{P1}b_\mathrm{P2}^{\dagger}b_\mathrm{X2}b_\mathrm{S}^{\dagger}\right)+\mathrm{H.c.},
\end{align}
in which $\Lambda$ is a constant quantifying the strength of the nonlinear interaction, and $b_\mathrm{J}$ is the annihilation operator for mode J in the ring. Here we are interested in the SP-SFWM and BS-FWM terms, which lead to the generation or destruction of a single photon in the mode S, and thus are a source of noise in the generation of degenerate squeezing. In a single ring resonator, when the resonances are nearly equally spaced in frequency, these processes are an intrinsic source of noise, and their mitigation, for example by a properly selected pump detuning \cite{zhao2020near}, is typically associated with a weakening of the desired nonlinear process of DP-SFWM as well.

The photonic molecule approach described in the main text aims at suppressing these parasitic processes by engineering the electromagnetic field enhancement in the structure, which is directly related to the process efficiency. This is done by engineering the resonance positions, in analogy to modifying the energy levels of a true molecule, by coupling a second resonator to the ring used for squeezing. This allows to achieve the splitting of the X1 and X2 resonances, detuning them well away from their original frequencies, and thereby suppressing the above-mentioned undesirable parasitic processes of SP-SFWM and BS-FWM.  

One can estimate the efficacy of this strategy by considering the nonlinear Hamiltonian in terms of asymptotic fields \cite{liscidini2012asymptotic}, which are directly connected to the resonant electromagnetic field enhancement factors $\mathcal{F}(k)$. In this case, the generic four-photon Hamiltonian associated with any of DP-SFWM, SP-SFWM, and BS-FWM can be written as:

\begin{equation}
    H_{\mathrm{NL}}=-\int\mathrm{d}k_{1}\mathrm{d}k_{2}\mathrm{d}k_{3}\mathrm{d}k_{4}S\left(k_{1},k_{2},k_{3},k_{4}\right)a_{k_{1}}^{\dagger}a_{k_{2}}^{\dagger}a_{k_{3}}a_{k_{4}}+\mathrm{H.c.},
\end{equation}
where $k$ is the wavevector, $a_k$ the annihilation operator of the (continuum) field associated with $k$, and 

\begin{equation}\label{Eq. S_asy_states}
    S(k_{1},k_{2},k_{3},k_{4})=\frac{3\Omega\bar{\Gamma}^{(3)}}{8\epsilon_{0}\mathcal{A}}\sqrt{\frac{\hbar\omega_{k_{1}}\hbar\omega_{k_{2}}\hbar\omega_{k_{3}}\hbar\omega_{k_{4}}}{(2\pi)^{4}}}\mathcal{F}(k_{1})\mathcal{F}(k_{2})\mathcal{F}(k_{3})\mathcal{F}(k_{4}),
\end{equation}
where $\mathcal{A}$ is the effective area, $\Omega$ a proportionality constant, $\bar{\Gamma}^{(3)}$ the third-order susceptibility \cite{bhat2006hamiltonian}, and $\mathcal{F}(k_i)$ the field enhancement factor at $k_i=\omega_i n_{\mathrm{eff}}(\omega_i)/c$, with $n_{eff}(\omega_i)$ the effective index and $c$ the speed of light. 

It is useful to define the overall resonant field enhancement associated with the desired four-wave process  
\begin{equation}
\mathbf{F}(\omega_1,\omega_2,\omega_3,\omega_4)=\mathcal{F}(k(\omega_1))\mathcal{F}(k(\omega_2))\mathcal{F}(k(\omega_3))\mathcal{F}(k(\omega_4)),
\end{equation}
which allows one to compare the strength of the various Hamiltonian terms for different structures and pumping configurations. 

\begin{figure}
    \centering
    \includegraphics[width=0.7\textwidth]{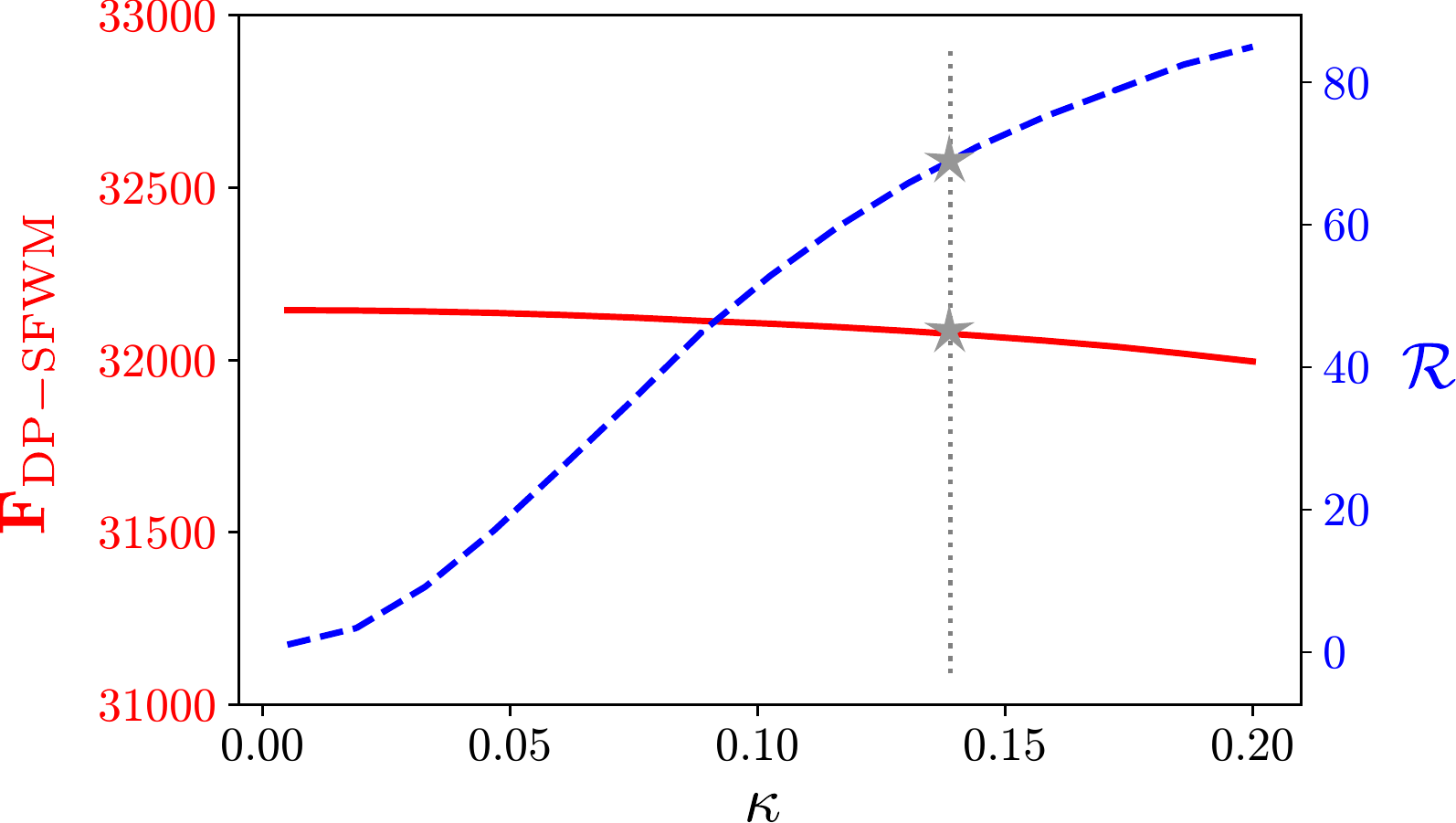}
    \caption{Field enhancement $\mathbf{F}_\mathrm{DP-SFWM}$ associated with the desired DP-SFWM process in the photonic molecule structure (red solid line, left vertical axis), and relative noise suppression ratio $\mathcal{R}$ (blue dashed line, right vertical axis) as a function of the coupling $\kappa$ between the principal and auxiliary ring. The dashed grey line indicates the approximate coupling chosen for the device reported in the main text; grey star markers indicate the operational point obtained for field enhancement and relative suppression.}
    \label{fig:field-enhancement}
\end{figure}

In Fig.~\ref{fig:field-enhancement} we plot $\mathbf{F}_\mathrm{DP-SFWM}(\omega_\mathrm{S},\omega_\mathrm{S},\omega_\mathrm{P1},\omega_\mathrm{P2})$ for DP-SFWM in our photonic molecule as a function of the field coupling coefficient $\kappa$ between the generating ring and the auxiliary ring necessary to split the resonances at $\omega_\mathrm{X1}$ and $\omega_\mathrm{X2}$ associated with the parasitic processes. Here  $\omega_\mathrm{P1}$ and $\omega_\mathrm{P2}$ are the pump frequencies and $\omega_\mathrm{S}=(\omega_\mathrm{P1}+\omega_\mathrm{P2})/2$ is that of the generated photons. The system has been designed such that all the three fields are resonantly coupled to the generating ring. As expected, the resonances at $\omega_\mathrm{P1}$, $\omega_\mathrm{P2}$, and $\omega_\mathrm{S}$ are only very weakly dependent on $\kappa$ in position and quality factor, thus $\mathbf{F}_\mathrm{DP-SFWM}(\omega_\mathrm{S},\omega_\mathrm{S},\omega_\mathrm{P1},\omega_\mathrm{P2})$ remains nearly constant.

In Fig.~\ref{fig:field-enhancement} we plot also the ratio 
\begin{equation}
    \mathcal{R}=\frac{\mathbf{F}_\mathrm{DP-SFWM}(\omega_\mathrm{S},\omega_\mathrm{S},\omega_\mathrm{P1},\omega_\mathrm{P2})}{\mathbf{F}_\mathrm{SP-SFWM}(\omega_\mathrm{S},\omega_\mathrm{X1},\omega_\mathrm{P1},\omega_\mathrm{P1})},
\end{equation}
which allows one to estimate the relative suppression of the parasitic process associated with the generation of one photon at $\omega_\mathrm{S}$ and one photon at $\omega_\mathrm{X1}$ due to SP-SFWM associated with the pump at  $\omega_\mathrm{P1}$. A similar result could be shown for SP-SFWM involving the resonance at  $\omega_\mathrm{X2}$ or BS-FWM. One can see that as a result of the splitting of the resonance at $\omega_\mathrm{X1}$ the overall field enhancement for the parasitic process decreases with $\kappa$ and is attenuated by about a factor of 60 (relative to the enhancement maintained by the desired DP-SFWM process) at our operating point. Achieving the same result by symmetrically detuning the pumps from their respective resonances (the approach taken by Zhao et al. \cite{zhao2020near}) would be possible, but it would also result in decreasing the desired field enhancement $\mathbf{F}_\mathrm{DP-SFWM}(\omega_\mathrm{S},\omega_\mathrm{S},\omega_\mathrm{P1},\omega_\mathrm{P2})$ by three orders of magnitude compared to the value obtained with our photonic molecule.

\add{It is also important to consider whether other four-wave mixing processes, involving resonances outside those considered in this discussion, could contribute noise to the squeezing band. The leading contribution from such processes involves cascaded stimulated and spontaneous four-wave mixing: classical stimulated four-wave mixing leads to the generation of coherent light at a pair of resonances with frequencies three free spectral ranges above and below the S resonance. These stimulated waves are weak compared to the pumps, having at most a few mW of generated light for the pump powers used, but they are strong enough to warrant concern over the corresponding spontaneous four-wave mixing processes that could add noise photons to the squeezing band.}

\add{The single-pump processes involving light generated in the stimulated modes are suppressed by the auxiliary resonator. The ring radius of that resonator is chosen to hybridize every fourth resonance of the principal resonator, and it accomplishes this over a very wide span of wavelengths. Thus the resonances lying six free spectral ranges above and below the S resonance are also strongly hybridized (i.e., split and detuned). These resonances are involved in the single-pump spontaneous processes that would add photons to the S mode via SP-SFWM arising from the stimulated waves acting as pumps. Therefore such processes are similarly suppressed in our device.}

\add{However, dual-pump processes involving spontaneous four-wave mixing driven by one of the bright pumps, and one of the stimulated waves, is not suppressed by the auxiliary resonator. These processes are much weaker than the other spontaneous processes discussed so far, since they involve one of the weaker, stimulated waves acting as a pump; they are automatically suppressed by approximately the ratio between the power in the generated stimulated waves and the power in the input pumps. In our experiments, this ratio is on the order of $10^{-3}$ for the pump powers that optimize squeezing, and thus about $30$ dB of suppression for this process is automatically ensured. This could be improved by the addition of extra rings specifically designed to suppress stimulated processes.}

\add{A complete accounting of every four-wave mixing process possible in this system is beyond the scope of this work. A full discussion and analysis is left to future studies.}

\newpage
\add{\section{Theory of degenerate squeezing}}
\add{To model the generation of degenerate squeezing in a microring resonator system for which noise suppression has been implemented, we model the intra-resonator dynamics using a set of coupled mode equations for the two pump resonances and the S resonance. We follow the same treatment used for calculating nondegenerate quadrature squeezing in the Supplementary Materials section S1 of Vaidya \textit{et al.} \cite{vaidya2020broadband}; here we present a summary of this calculation adapted to the degenerate case. }

\add{Treating the pumps classically, we represent their steady state intra-resonator amplitudes as $\overline{\beta}_\mathrm{P1}e^{-i(\omega_\mathrm{P1}-\Delta_\mathrm{P1})t}$ and $\overline{\beta}_\mathrm{P2}e^{-i(\omega_\mathrm{P2}-\Delta_\mathrm{P2})t}$, where $\Delta_\mathrm{P1}$ and $\Delta_\mathrm{P2}$ are the detunings of each pump from resonance, and $\overline{\beta}_\mathrm{P1}$ and $\overline{\beta}_\mathrm{P2}$ are constants. The equation of motion for the S resonance mode is then
\begin{equation}\label{eqn:b_s_dynamics}
    \left( \frac{d}{dt} + \overline{\Gamma}_\mathrm{S} + i\omega_\mathrm{S} - 2i\Lambda(|\beta_\mathrm{P1}|^2 + |\beta_\mathrm{P2}|^2)\right)b_\mathrm{S}(t) = -i\gamma_\mathrm{S}^*\psi_{\mathrm{S}<}(0,t) - i\mu_\mathrm{S}^*\phi_{\mathrm{S}<}(0,t) + 2i\Lambda b_\mathrm{S}^\dagger(t) \overline{\beta}_\mathrm{P1}\overline{\beta}_\mathrm{P2}e^{-i(\omega_\mathrm{P1}+\omega_\mathrm{P2}-\Delta_\mathrm{P1} - \Delta_\mathrm{P2})t},
\end{equation}
where $\overline{\Gamma}_\mathrm{S}$ is the damping rate of the S resonance, related to the loaded quality factor $Q_\mathrm{S}$ via $Q_\mathrm{S}=\omega_\mathrm{S}/2\overline{\Gamma}_\mathrm{S}$, and $\gamma_{\mathrm{S}}$ and $\mu_\mathrm{S}$ are the coupling coefficients to the input channel vacuum field and loss channel vacuum field $\psi_{\mathrm{S}<}(0,t)$ and $\phi_{\mathrm{S}<}(0,t)$, respectively.
}

\add{The squeezing spectrum $S(\Omega)$ can be calculated via the relation
\begin{equation}
    S(\Omega)= 1 + N(\Omega,\Omega) + N(-\Omega,-\Omega) + 2\mathrm{Re}\lbrace e^{-2i\phi_\mathrm{LO}}M(\Omega,-\Omega)\rbrace,
\end{equation}
where $\phi_\mathrm{LO}$ is the local oscillator phase and the moments $M$ and $N$ are defined via
\begin{equation}
    M(\Omega,\Omega')\delta(\Omega-\Omega')=v_S\langle \psi_{\mathrm{S}>}(\Omega)\psi_{\mathrm{S}>}(\Omega')\rangle
\end{equation}
and
\begin{equation}
    N(\Omega,\Omega')\delta(\Omega-\Omega')=v_S\langle \psi_{\mathrm{S}>}^\dagger(\Omega)\psi_{\mathrm{S}>}(\Omega')\rangle,
\end{equation}
where $\psi_{\mathrm{S}>}(\Omega)$ is the output channel field at sideband frequency $\Omega$. Calculating these quantities from the solution obtained to Eq. \ref{eqn:b_s_dynamics}, we obtain
\begin{equation}\label{eqn:sqz_spec}
    S_\pm(\Omega) = 1 + \frac{4\eta g\left(2g\pm\sqrt{\left(\frac{\Omega^2}{\overline{\Gamma}_\mathrm{S}^2}+ 1+g^2-\frac{\Delta^2}{\overline{\Gamma}_\mathrm{S}^2}\right)^2 + 4\frac{\Delta^2}{\overline{\Gamma}_\mathrm{S}^2}}\right)}{\left(\frac{\Omega^2}{\overline{\Gamma}_\mathrm{S}^2}- 1+g^2-\frac{\Delta^2}{\overline{\Gamma}_\mathrm{S}^2}\right)^2 + 4\frac{\Omega^2}{\overline{\Gamma}_\mathrm{S}^2}},
\end{equation}
where $S_+(\Omega)$ and $S_-(\Omega)$ are respectively the maximum (over all quadrature angles) anti-squeezing and squeezing at sideband $\Omega$, $\Delta$ is the overall net detuning associated with the degenerate spontaneous four-wave mixing process, $\Delta=2\Lambda(|\beta_\mathrm{P1}|^2+|\beta_\mathrm{P2}|^2)-(\Delta_\mathrm{P1}+\Delta_\mathrm{P2})/2$ including both pump detuning and XPM-induced detuning, $\eta$ is the overall total system transmission efficiency including the ring escape efficiency and collection and detection efficiencies, and $g$ is a dimensionless gain parameter defined as
\begin{equation}\label{eqn:g_defn}
    g=\frac{2\Lambda|\beta_\mathrm{P1}\beta_\mathrm{P2}|}{\overline{\Gamma}_\mathrm{S}}.
\end{equation}
In this treatment, the OPO threshold corresponds to a gain parameter $g=1$. }

\add{
The expression \ref{eqn:sqz_spec} is used to fit the experimental squeezing spectrum traces in Figs. \ref{fig:results}(c-d) of the main text. The independently measured parameters for $\eta = 0.34$, $\overline{\Gamma}_\mathrm{S}=\omega_\mathrm{S}/2Q_\mathrm{S}\approx 2.0\times 10^9\mathrm{s}^{-1}$, and $\Delta=0.15\times\overline{\Gamma}_\mathrm{S}$ were used to fit \ref{fig:results}(d), with $g$ kept as the free fitting parameter. The resulting best least-squares fit yielded $g\approx 0.46$, consistent with our estimate of the corresponding pump power being about half that of the OPO threshold. This fit was performed simultaneously to the squeezing and anti-squeezing data, i.e., the parameter $g$ was constrained to be equal for both the maximum and minimum quadrature variance data across the range of sidebands in the spectrum. The fit for the data in Fig. \ref{fig:field-enhancement}(c) used the same parameters and $\Omega=2\pi\times 20$ MHz, and assumed a linear relationship between the total pump power and $g$, consistent with the definition Eq. \ref{eqn:g_defn}. The proportionality coefficient between $g$ and the input power was kept as the free fitting parameter. With this form, the fit predicted the gain parameter at the pump power that optimized squeezing to be $g\approx 0.50$, again consistent with our estimate of the OPO threshold and the result of the fit to the squeezing spectra in Fig. \ref{fig:results}(d). The three data points at the highest powers were omitted from the fit in Fig. \ref{fig:results}(c), as they are significantly affected by phase noise, which is not included in the model. Inclusion of these points does not significantly change the extracted fit parameters. Finally, it should be noted that the exact OPO threshold depends on other factors not included in this model, such as dispersion which would give rise to different net effective detunings. The consistency between the extracted fit parameters and the estimated OPO threshold is therefore not an especially precise statement; that the fit parameters independently extracted from the two fits are consistent is a more significant validation of our model. 
}
\end{document}